\documentclass[preprint,review,12pt]{elsarticle}

\usepackage{graphicx}

\usepackage{mathptmx}                          
\usepackage{amsmath,amssymb}
 \usepackage{amsthm}
\usepackage{booktabs}                          
\usepackage{soul}
\usepackage{lineno}

\newcommand{\wea}{WH}

\journal{Planetary and SPace Science}

\begin{document}

\begin{frontmatter}



\title{Fugitives from the Hungaria Region: close encounters and impacts with terrestrial planets.}


\author[a]{M. A. Galiazzo} 
\author[a]{\'A. Bazs\'o}
\author[a]{ R. Dvorak}

\address[a]{Institute of Astronomy, University of Vienna,
T{\"u}rkenschanzstrasse 17, A-1180 Wien, Austria\\  Email: mattia.galiazzo@univie.ac.at}


\begin{abstract}
Hungaria asteroids, whose orbits occupy the region in element space between $1.78< a< 2.03$ AU, 
$e<0.19$, $12^\circ<i<31^\circ$, are a possible source of Near-Earth Asteroids (NEAs).
Named after 
(434) Hungaria these asteroids are
relatively small, since the largest member of the group has a diameter of just about 11 km.
 They are mainly perturbed by Jupiter and 
Mars, possibly becoming Mars-crossers and, later, they may even cross the orbits of Earth and Venus. 
In this paper we analyze the close encounters and possible impacts of escaped Hungarias with the terrestrial planets.
Out of about 8000 known Hungarias we selected 200 objects which are on the edge of the group.
We integrated their orbits over 100 million years in a simplified model of the planetary system (Mars to Saturn)
 subject only to gravitational forces.
We picked out a sample of 11 objects (each with 50 clones) with large
variations in semi-major axis and restarted the numerical integration in a gravitational
model including the planets from Venus to Saturn.
 Due to close encounters, some of them achieve high inclinations and
eccentricities which, in turn, lead to relatively high velocity impacts on Venus,
Earth, and Mars.
 We statistically analyze all close encounters and impacts with the
terrestrial planets and  determine the encounter and impact velocities of these fictitious Hungarias.
\end{abstract}


\end{frontmatter}


\section{Introduction}

For the purposes of this paper we consider as belonging to the Hungaria group 
the asteroids whose orbital elements are in the range $1.78 (\mathrm{AU}) < a < 2.03 (\mathrm{AU})$, 
$12^{\circ} < i < 31^{\circ}$ and $e<0.19$, so that our sample finally consists of 8258 asteroids; the orbital data were taken from the ASTORB database (http://www2.lowell.edu/elgb, updated August 2010).
This orbital region is confined by the $\nu_6$ secular resonance and by the mean motion resonances 
4:1 with Jupiter and 3:4 with Mars. \\

There is evidence that some meteorites may originate from the Hungarias. This is deduced  in the first place from the spectra of the majority of the Hungarias; about $77\%$ of them are of the E-type, $17\%$ belong to the S-type and $6\%$ to the C-type \citep{War2009}.
The E-type asteroids are consistent with the composition of some meteorites (aubrites, \citet{Zel1977}) found on the Earth.

 Because of their typical composition, the Hungarias have an average albedo of $p_v \sim 0.4$ (\citet{War2009}, hereafter abbreviated as \wea), which distinguishes them from other asteroids in the main belt having $p_V \sim 0.18$ or lower.

The largest member of the group, (434) Hungaria, has a diameter of just about 11 km \citep{Mor1977}.

\label{origins}
The majority of Hungarias have a retrograde rotation and similar spin rates \citep{Pra2008, Ros2009}. \citet{War2007} found a consistent group of binaries (more than $10\%$) with fast rotating primaries, this presence being a sign of a collisional origin \citep{Dur2004, Zap2002}. Based on the study of \citet{Lem1994} on proper elements \wea{} assumed that the Hungarias formed after a catastrophic collision of (434) Hungaria, presumably the largest fragment of the Hungaria collisional family. Starting from this collisionary assumption they computed an age for the family of about 0.5 Gy, that comes from the degree of spreading versus size of family members. \wea{}, considering 2859 family members,
found a value of 26 km for the diameter of the putative parent body.

\citet{Mil2010} confirmed this collisional origin, underlining the possibility
of the presence of a subfamily, especially for the uniform number distribution
in semi-major axis for values above 1.92 AU. They suggested a half life of 960 My and a diameter of 30 km for the parent body, this latter value in good agreement with\wea{}. \citet{Bot2011}, in contrast to the previous suggestions about the origin, assumed that the Hungarias evolved from the depletion of a part of the primordial main belt with semi-major axes between 1.7 AU and 2.1 AU.

In the rest of this paper we present the results of our statistical investigation of the close encounters of the escaped Hungarias with the terrestrial planets, in a model including only gravitational interactions. The methods used are described in section 2; the results are shown in section 3 and in section 4 we discuss the production of NEAs from the Hungaria population. The conclusions are in section 5.

\section{Methods}
\label{enlarge}

Out of our 8258 Hungarias, we selected 200 according to a criterion based on the action variables (semi-major axis, $a$, eccentricity, $e$ and inclination, $i$ of the asteroid); we chose the following variable:\\
\begin{displaymath}
d= \sqrt{(\frac{e}{<e>})^2 + (\frac{a}{<a>})^2 +(\frac{\sin{i}}{<\sin{i}>})^2}
\end{displaymath}
 and picked up the 200 Hungarias with the highest values of $d$.\\
 Our study of the dynamical transport to the terrestrial planets and the possible impacts was performed in
two steps:

\begin{enumerate}

\item 
We integrated the orbits of the 200 asteroids in a simplified dynamical model for
the solar system (Sun, Mars, Jupiter, Saturn and the massless
asteroids) for 100 million years (My) to identify possible escapers (that from now on we call ``fugitives''), based on
the variation in of the semi-major-axis; in this way we identify 11 fugitives.

\item
The 11 fugitives found (see Table \ref{escapers}) were then cloned, adding to each of them 49 additional sets of initial conditions (see later). We integrated the 550 objects in a model Solar System including the planets from Venus to Saturn, again for 100 My; all the results presented in the paper referred to the outputs of these integrations.

\end{enumerate}

All the integrations have been done using the Lie-series integrator \citep{Egg2010, Han1984}.

The criterion of selection for the fugitives was to check whether the excursion in semi-major axis $\Delta a=a_{max}-a_{min}$ for an individual object was $\Delta a > \Delta a_{group}/16=0.0156$ AU, where $\Delta a_{group}$ corresponds to $\sim7\%$ of the total range of semi-major axis spanned by the Hungarias. 

For the generation of clones, we generated random values for $(a,e,i)$, starting with the initial conditions of the
 escapers in the following ranges: $a \pm 0.005$ AU, $e \pm 0.003$ and $i \pm 0.005^\circ$.

It turns out that the 11 fugitives belong to the Hungaria family (see the AstDyS website, http://hamilton.dm.unipi/astdys/, for comparisons with the elements), so we can speak about fugitives from the Hungaria \emph{family}. Two of them (30935) Davasobel, and (211279) 2002 RN$_{137}$ are also present in the list of \cite{Mil2010} for the Hungaria asteroids in strongly chaotic orbits.

All the data relative to the close encounters with the terrestrial planets were stored for later examination; for the Earth we took as close encounter limiting distance the average lunar distance, $2.50\cdot10^{-2}$ AU, for Mars and Venus, we used a distance scaled in proportion to the ratio of their Hill spheres with respect to the one of the Earth, namely $1.70\cdot10^{-2}$ AU for Venus and $1.66\cdot10^{-2}$ AU for Mars.

\begin{table}
\begin{center}
\begin{tabular}{lccc}
\toprule
Asteroid     & $a$ [AU] & $e$ & $i$ [deg] \\
\midrule
 (211279) 2002 RN$_{137}$ & 1.8538 & 0.1189 & 22.82\\
  (41898) 2000 WN$_{124}$ & 1.9073 & 0.1062 & 17.11\\
  (39561) 1992 QA         & 1.8697 & 0.1116 & 26.23\\
  (30935) Davasobel       & 1.9034 & 0.1178 & 27.81\\
 (171621) 2000 CR$_{58}$  & 1.9328 & 0.1051 & 17.19\\
 (152648) 1997 UL$_{20}$  & 1.9894 & 0.1841 & 28.88\\
 (141096) 2001 XB$_{48}$  & 1.9975 & 0.1055 & 12.32\\
  (24883) 1996 VG$_9$     & 1.8765 & 0.1556 & 22.71\\
  (41577) 2000 SV$_2$     & 1.8534 & 0.1843 & 24.97\\
 (129450) 1991 JM         & 1.8512 & 0.1263 & 24.50\\
 (175851) 1999 UF$_5$     & 1.9065 & 0.1874 & 19.24\\
\bottomrule
\end{tabular}
\caption{Osculating elements for the escaping Hungarias: semi-major axis ($a$), eccentricity ($e$), inclination ($i$) in degrees.}
\label{escapers}
\end{center}
\end{table}

\noindent


\section{RESULTS}

\label{uno}

Once our fugitives have left the Hungaria region, their orbital evolution becomes strongly chaotic
 and is strongly affected by close encounters with one or more of the terrestrial planets.
 
(211279) 2002 RN$_{137}$ is one example for these Hungarias fugitives that becomes NEA: it changes from one NEA-type to 
another, starting to be an Amor at 64.015 My and then an Apollo for the first time at 72.669 My.\\

We  found  that, when the fugitives start to be NEAs, some observed NEAs pass through
 similar semi-major axis and eccentricities. In fact, 
more than 70\% of the clones are inside these ranges: $1.71 (\mathrm{AU})<a<2.04 (\mathrm{AU})$ and $0.21<e<0.38$, with a peak around
$1.92 (\mathrm{AU})<a<2.04 (\mathrm{AU})$ and $0.32<e<0.38$; inside the first range we have $\sim 880$ known NEAs, while inside the 
second one we find 61 known NEAs (from http://ssd.jpl.nasa.gov/sbdb\_query.cgi\#x, JPL Small-Body Database 
Search Engine). 
In particular we mention 143409 (2003 BQ$_{46}$), 249595 (1997 GH$_{28}$) and 
285625 (2000 RD$_{34}$), because they are inside the range of the absolute magnitude in visual ($H_V$) of
 Hungarias,  $14<H_v<18$ (for details see Fig.3 of \citep{War2009}).


\subsection{Close encounters}

The sample of Hungarias which we used has a high probability ($>77\%$) to have close
encounters (CEs) with terrestrial planets, especially with Mars, in the 
100 My time interval. We show the detailed results in 
Tab.~\ref{single} counting the CEs for every planet separately (first three
columns) and the CEs (last three columns) exclusively to only one of the terrestrial planets.

Not surprisingly, most of the fictitious asteroids have CEs with Mars. The
CEs to Earth and Venus tend to take place in the expected order, because the dynamical
transport the inner system needs more time for Venus than for the Earth
and Mars. Almost half of our bodies just meet Mars and then undergo either an ejection or a collision
 with the Sun (see later). None of them have only CEs with the Earth and only one of them has CEs
 exclusively with Venus and the Earth; 3 of them have CEs exclusively with Mars and Venus.

In the first three columns of Tab.~\ref{pairs} the encounter probabilities for two
of the planets are not so different for any pair and also not for CEs with all
the three planets. The possibility that an object meets only a specific pair is very low
especially for Mars-Venus and Earth-Venus, but it is more probable for Mars-Earth.


\begin{table}
  \begin{center}
  \begin{tabular}{lcccccc}
    \toprule
           & \multicolumn{3}{c}{close encounters to \ldots} &
\multicolumn{3}{c}{close encounters only to \ldots}\\
    \cmidrule(r){2-4} \cmidrule(r){5-7}
           & Mars  & Earth & Venus & Mars  & Earth & Venus\\
    \midrule
N.     & 427    & 158    & 132    & 243  & 0   & 0\\
\% & 77.6  & 28.7  & 24.0  & 44.2  & 0.0  & 0.0\\
    \bottomrule
  \end{tabular}
  \caption{Close encounters for the sample of 550 Hungarias during 100 My
           with the terrestrial planets. Columnn 2 to 4 for every planet
           separately, Columns 5 to 7 for a planet alone. ``N.'' stands for
           number and ``\%'' for percentage.}
  \label{single}
  \end{center}
\end{table}


\begin{table}
  \begin{center}
  \begin{tabular}{lccccccc}
    \toprule
           & \multicolumn{4}{c}{close encounters to pair \ldots} &
\multicolumn{3}{c}{close encounters only to pair \ldots}\\
    \cmidrule(r){2-5} \cmidrule(r){6-8}
           & M \& E & M \& V & E \& V & M \& E \& V & M \& E & M \& V & E \& V\\
    \midrule
N.     & 143    & 122    & 119    & 138 & 24 & 3   & 1\\
\% & 26.0   & 22.2   & 21.6   & 25.1 & 4.4& 0.6    & 0.2\\
    \bottomrule
  \end{tabular}
  \caption{Close encounters for the sample of 550 Hungarias during 100 My
           with the terrestrial planets. Columnn 2 to 4 for at least two
           planets and columns 5 to 7 ONLY for two planets. M stands for Mars, E for
           the Earth and V for Venus. ``N.'' stands for
           number and ``\%'' for percentage.} 
  \label{pairs}
  \end{center}
\end{table}

In Tab.~\ref{CE-time} one can see that the average time for a first close
encounter with Venus is quite long ($\sim 63$ My), but it is somewhat surprising
 that this time is not much smaller for the Earth ($\sim 62$ My).
 For the first CE with Mars the average time is very low ($\sim 14$ My);
 we already explained it with the relative closeness of the semi-major axes
 of Mars with the perihelion distances of the Hungarias. 

In Tab.~\ref{Family} the average time $\langle T_{pl} \rangle$ for an object 
to become member of a NEA type (i.e. Amor, Apollo, Aten and IEO) is listed.  These four
groups are defined as follows:

\begin{itemize}

\item {\bf IEOs (Inner Earth Orbits)\footnote{Sometimes we will call them also Atiras.}}  
	NEAs whose orbits are contained entirely with the orbit of the
  Earth with   	$a<1.0$ AU,
  $Q<0.983$ AU

\item {\bf Atens} 
Earth-crossing NEAs with semi-major axes smaller than Earth's, named after (2062) Aten -- with $a<1.0$ AU, 
$Q>0.983$ AU

\item {\bf Apollos}
Earth-crossing NEAs with semi-major axes larger than Earth's, named after (1862) Apollo, with $a>1.0$ AU, 
$q<1.017$ AU

\item {\bf Amors}
Earth-approaching NEAs with orbits exterior to Earth's but interior to Mars, named after (1221) Amor, with  	
$a>1.0$ AU, $1.017 (\mathrm{AU})<q<1.3 (\mathrm{AU})$ 

\end{itemize}

The Table shows how $\langle T_{pl} \rangle$ increases going from the outer NEAs (the Amors) to the inner 
ones (the IEOs).

\begin{table}
\begin{center}
\begin{tabular}{lcc}
\toprule
  Planet                    &  $\langle T_{pl} \rangle$ [My] & $T_{pl,\mathrm{min}}$ [My]\\
\midrule
  Venus                     &  63.34                         &  (41577) 2000 SV$_2$ 10.9\\
  Earth                     &  62.45                         & (152648) 1997 UL$_{20}$ 10.5\\
  Mars                      &   13.51                        &  (24883) 1996 VG$_9$ 0.07\\
\bottomrule
\end{tabular}
\caption{Average value of the time of the first CE with the planets for all
  objects ($2^{nd}$column) and for the shortest time of a first CE with the planets for a  single object $3^{rd}$column.}
\label{CE-time}
\end{center}
\end{table}

\begin{table}
\begin{center}
\begin{tabular}{lcc}
\toprule
Family   & $\langle T \rangle$ [My] & $T_\mathrm{min}$ [My] \\
\midrule
  Amor   & 46.67 & (41577) 2000 SV$_2$ 0.09   \\
  Apollo & 60.82 & (152648) 1997 UL$_{20}$ 1.68  \\
  Aten   & 61.44 & (152648) 1997 UL$_{20}$ 8.40 \\
  Atira  & 69.37 & (152648) 1997 UL$_{20}$ 8.92 \\
\bottomrule
\end{tabular}
\caption{$T$ is the average time (of the real asteroid and the clones) when the object 
becomes member of the NEA family, $T_\mathrm{min}$ is the minimum time for the indicated object.}
\label{Family}
\end{center}
\end{table}

\begin{table}
\begin{center}
\begin{tabular}{lcccc}
\toprule
  Asteroid    & Imp. Sun $\%/100My$ & $\langle P  \rangle$ [AU/100My]
&$q$ [AU]\\
\midrule
 (211279) 2002 RN$_{137}$ &  4 & $0.90 \pm 0.33$ & 0.00\\
 (41898) 2000 WN$_{124}$  &  0 & $1.55 \pm 0.07$ & 1.02\\
  (39561) 1992 QA         & 22 & $0.96 \pm 0.30$ & 0.00\\
  (30935) Davasobel       & 14 & $0.69 \pm 0.19$ & 0.00\\
 (171621) 2000 CR$_{58}$  &  2 & $1.61 \pm 0.17$ & 0.00\\
 (152648) 1997 UL$_{20}$  & 36 & $0.82 \pm 0.34$ & 0.00\\
 (141096) 2001 XB$_{48}$  & 50 & $0.54 \pm 0.23$ & 0.00\\
  (24883) 1996 VG$_9$     &  4 & $1.37 \pm 0.07$ & 0.00\\
  (41577) 2000 SV$_2$     & 40 & $0.68 \pm 0.37$ & 0.00\\
 (129450) 1991 JM         &  4 & $1.20 \pm 0.20$ & 0.00\\
 (175851) 1999 UF$_5$     & 20 & $1.07 \pm 0.28$ & 0.00\\
\bottomrule
\end{tabular}
\caption{Number of clones per 100 My per asteroid that end up as
sun-impactors (we assume an impact with the Sun at whenever its perihelion is less than 0.0047 AU,
 approximately the radius of the Sun) with the average and the minimum perihelion of their clones.}
\label{sun}
\end{center}
\end{table}

\begin{table}
\begin{center}
\begin{tabular}{lcccccc}
\toprule
   Family/group     & T [My] & a [AU] & e &
i [deg] & T. Par.& Var. in \%\\
\midrule
   \multicolumn{5}{l}{(211279) 2002 RN$_{137}$}\\
\midrule
   Hungaria         & 0          & 1.8552  & 0.1175    & 22.82 &
1.51625& 0.0\\
   First enc. Mars  & 0.050      & 1.8551  & 0.1625    & 25.46 &
1.48297& 2.2\\
   leaves Hungarias & 0.063      & 1.8540  & 0.1904    & 23.47 &
1.49585& 1.4\\
   NEA - Amor       & 64.016     & 1.7817  & 0.2719    & 29.09 &
1.40315& 7.5\\
   NEA - Apollo     & 72.670     & 1.8166  & 0.4478    & 28.23 &
1.33699&11.9\\
\midrule
   \multicolumn{5}{l}{(41577) 2000 SV$_2$}\\
\midrule
   Hungaria         & 0          & 1.8517  & 0.1850    & 24.97 &
1.48231& 0.0\\
   leaves Hungarias & 0.004      & 1.8517  & 0.1988    & 24.16 &
1.48684& 0.3\\
   NEA - Amor       & 0.085      & 1.8490  & 0.2991    & 24.29 &
1.45306& 2.0\\
   First enc. Mars  & 0.226      & 1.8501  & 0.2331    & 23.25 &
1.48553& 0.2\\
   First enc. Earth & 77.926     & 2.0553  & 0.5048    &  3.98 &
1.47785& 0.3\\  
   First enc. Venus & 78.544     & 2.1634  & 0.7259    &  5.61 &
1.23792& 16.5\\  
   NEA - Apollo     & 77.889     & 2.0485  & 0.5036    &  0.73 &
1.48050& 0.1\\
   Sun-impact       & 85.545     & 2.3541  & 0.9999    & 20.89 &
0.23267& 84.3\\
\bottomrule
\end{tabular}
\caption{Dynamical evolution for clones (211279) 2002 RN$_{137}$ and (41577) 2000 SV$_2$
describing to which asteroid group the objects belong to; in addition the
semi-major axes a, the eccentricity e, the inclination i and the Tisserand
parameter and its variation are given in columns 3 to 6.}
\label{membership}
\end{center}
\end{table}

In Tab.~\ref{sun} we show
the percentage of the clones which collide with the Sun. 
In particular the clones of
(41577) 2000 SV$_2$ have the highest probability to hit the Sun, while also (152648) 1997 UL$_{20}$, 
(141096) 2001 XB$_{48}$ and (39561) 1992 QA have high chances to do so (Table \ref{sun}). 

According to our statistics we can say
that about $18$ \% from the escaping Hungarias end up as Sun colliders within 100 My. 

 From the first sample (200 fictitious
asteroids out of $\sim$ 8000)  we found only 5\% of escapers, so our estimate of an upper limit of 
Sun-impactor Hungarias
in the time scale of 100 My is $ 0.98\%$, instead for the planet's impacts ($ 0.23\%$). Only a very
minor fraction escape beyond Jupiter's orbit and become Centaurs and afterwards  Trans-Neptunian Objects 
(TNOs), just $<1$\% of them.


\subsection{Analysis of close encounter data}

As an example of the dynamical evolution we discuss two examples out of
our sample, (211279) 2002 RN$_{137}$, that ends as member of the Apollo group, and 
(41577) 2000 SV$_2$, that has an impact on the Sun. Both show the characteristic
behaviour of changing the group membership inside the NEAs. 
In Tab.~\ref{membership} details on the escape times from the Hungaria region of these two asteroids 
are given. Here ``escapers'' mean that the Hungarias' orbital elements
($a,e,i$) fall outside the range of semi-major axes $a$, eccentricity $e$ ,or
inclination $i$ defined in 
section \ref{enlarge}.

In general, the fugitives leave the group very soon, sometimes in less than $10^5$ years, 
apart from (41898) 2000 WN$_{124}$, that has only 5 clones which escape after CEs to Mars.
If the escapers have CEs with Mars, they also become NEAs (Table \ref{CE-time}, note
that here 
the ones that becomes soon NEAs have a higher probability than the others to
become 
sun grazers, e.g. (41577) 2000 SV$_2$, Table ~\ref{membership}) , in particular Apollos and Amors, but changing
orbital type 
during the last millions of years of the integration.
\citet{Sho1979} showed, that asteroids frequently change the NEA 
family due to their chaotic orbits.

 When belonging to the Amor group within 11 My (average on the whole) or less,
 they 
become Apollos, and in many cases also Atens and Atiras, e.g. (152648) 1997 UL$_{20}$, whose clones have a 
huge number of close encounters with all the terrestrial planets.
 Their eccentricities will raise up after close encounters with Mars, and 
then impacts on all terrestrial planets are possible, even on the Sun.
Observing the average inclination in Table \ref{posch}, the inclination of the asteroids that
 have very likely become NEAs, like (211279) 2002 RN$_{137}$, (30935) Davasobel and (39561) 1992 QA,
 increases.
 
It is interesting to see that the normalised Tisserand parameter 
($T_{ast}=\frac{1}{2a}+\sqrt{a(1-e^2)}cos(i)$, where 
$a$ is the semi-major axis of the asteroid orbit, $e$ the eccentricity and
$i$ the inclination) stays almost constant after different close encounters also with the Earth:
 see Table ~\ref{membership}, where in the last column where we have put the variation in percentage of 
the first value. Till the asteroid is not a NEA, the parameter does not vary more than 2\% 
and stays close to the average value for the
Hungaria family, that is $1.546 \pm 0.017$ which, just for comparison to the
value for  the Vesta $1.740\pm0.006$ (from data by \cite{Gal2012}), is much lower.

From the close encounters we derive information about the duration (time while the body enters and 
exits from the region defined for a close encounter for each planet, see section 2) and 
the relative velocity in front of each terrestrial planets. This last one will be compared with the 
relative velocity of the observed NEAs.

We study the relation between the duration and the absolute number of close encounters, with one or 
more terrestrial planets. We underline that the mean duration of close encounters depends on how many planets
 the Hungaria-fugitive can approach: more often planets are reached, shorter are the duration of the encounters.

In fact for those objects that have the lowest probability to have close encounters we find a bigger duration,
 i.e. the maximum value with Mars is for (41898) 2000 WN$_{124}$ (0.65 d), which also has only a small
 probability to have a close approach with the planet, and does not approach the other two planets.

 Of course the singular event depends mainly on the entry velocity. The ``entry velocity'' is the velocity
 of the asteroid when it arrives for the first time at the maximum distance established for a close encounter from
 the planet.

 On average the mean duration of close encounters with Mars is about half a
 day, and this is the highest value for the terrestrial planets. In summary:
$0.27\pm0.05$ d for Venus, $0.36\pm0.09$ d for Earth and $0.55\pm0.05$ d for Mars. 

The Hungaria-fugitives seem to have more close encounters, even if the scattering of encounters' mean is relatively high,
 with the Earth respect to the other planets. In fact the average number of close encounters for each terrestrial planet
in 100 My of orbit-evolution is for Mars $57$, for the Earth $77$ and for Venus $42$, even if almost all of them has a
 close encounter with Mars, but not with the other two terrestrial planets (see Table \ref{single}).

The average deflection angle $\Theta$, i.e. the angle between the direction of the entry velocity vector
and the one of the exit velocity vector, is usually very small ($<1^\circ$), though in some cases we
get values of more than  $60^\circ$, and even more than
$90^\circ$ like for (129450) 1991 JM.

The mean entry velocities for the fugitives  are compared with the mean 
velocity of the real known asteroids with respect to each planet.

In order to compute the relative velocity of the known asteroids in front
 of the planets we considered
\begin{itemize}
\item as Venus-crossers: IEOs (the ones with $Q\geq0.718$ AU), Atens (the ones with $q\leq0.728$ AU)
 and Apollos (the ones with $q\leq0.728$ AU);
\item as Earth-crossers: Atens and Apollos;
\item as Mars-crossers: Mars crossing asteroids (the non-NEA ones with $1.300 (\mathrm{AU})<q<1.666 (\mathrm{AU})$ and $a<3.2$ AU), 
      and all the NEAs with $Q\geq1.381$ AU (so no IEOs, but a part of the other NEA-types).
\end{itemize}

The close encounter velocity of the real asteroids ($v_r$) is computed with the next equations:\\

$v_r=U v_{planet}$\\ 

$v_{planet}=\sqrt{\frac{G(M_{sun}+M_{planet})}{r_{planet}}}$\\ 

$U=\sqrt{3-T_{planet}}$\\

$T_{planet}=\frac{a_{planet}}{a_{ast}}+2\sqrt{\frac{a_{ast}}{a_{planet}}(1-e_{ast}^2)}\cos{i_{ast}}$\\

Where $G$, $M_{sun}$, $M_{planet}$ and $r_{planet}$ are respectively the 
gravitational constant, the mass of the Sun, the mass of the planet, and 
the heliocentric distance of the planet from the Sun, assumed 
as the semi-major axis of the planet, all values in SI, the international
system of measures. $U$ is the adimentional encounter velocity at infinitum computed via
the Tisserand parameter ($T_{planet}$) in respect to the relative planet. 
$a_{planet}$ is the semi-major axis of the relative planet,
 to which we compute the parameter, and the remnants values are the
 other osculatory elements of each asteroids to compare to the relative planet.

We take in account of the
 circular orbital velocity of each planet ($v_E=29.7892$, $v_M=24.1333
$ and $v_V=34.9084$, respectively for the
 one of the Earth, Mars and Venus, all in km/s).

\begin{table}
\begin{center}
\begin{tabular}{lccc}
\toprule
Sample & $<v_{r,Venus}>$   & $<v_{r,Earth}>$ & $<v_{r,Mars}>$       \\
\midrule
Hungaria fugitives &    $27.80\pm1.71$& $21.54\pm2.30$ & $10.85\pm1.01$        \\
\midrule
Real asteroids  & $21.66\pm8.69$&$15.18\pm7.35$&$10.83\pm4.99$         \\
\midrule
\bottomrule
\end{tabular}
\caption{Average velocity of the asteroids in  respective to Venus, the Earth, and Mars and standard deviation. The first row is the average of the entry velocity of the all Hungaria fugitives clones per planet, the second one is the average of the real asteroids that cross the orbit of the planets.} 
\label{relvel}
\end{center}
\end{table}

By our results the encounter velocity is greater for Hungarias than the 
mean, especially for Venus and the Earth, Mars is pretty the same,
 despite the high standard deviation. The values would give a more probable
 larger impact velocities in case of an impacts than the average
 one found for all type of impactors, so it seems that possible Hungarias' impactors
 are faster than the average.

\begin{table}
\begin{center}
\begin{tabular}{llccc}
\toprule
Planet   & Ast.       & $\bar{v}_{en}$ & $\bar{v}_{en,max}$ & $\bar{v}_{en,min}$ \\
\midrule
Venus 
         & (211279) 2002 RN$_{137}$  & $27.32\pm7.42$ & 35.10 & 19.26\\
\cmidrule{2-5}
         & (30935) Davasobel     & $29.76\pm7.38$ & 37.42 & 24.38 \\
\cmidrule{2-5}
         &(39561) 1992 QA  & $29.28\pm11.67$ & 42.03 & 22.66 \\
\midrule
Earth  
         & (211279) 2002 RN$_{137}$  & $20.03\pm6.64$ & 29.94 & 12.60 \\
\cmidrule{2-5}
         & (30935) Davasobel    & $20.36\pm6.07$ & 28.64 & 16.54 \\
\cmidrule{2-5}
         &(39561) 1992 QA    & $22.32\pm12.17$ & 31.32 & 13.05 \\
\midrule
Mars
         & (211279) 2002 RN$_{137}$   & $10.71\pm1.71$ & 16.63 & 5.96  \\
\cmidrule{2-5}
         & (30935) Davasobel    & $13.15\pm1.72$ & 19.12 & 7.91 \\
\cmidrule{2-5}
         & (39561) 1992 QA    & $11.05\pm1.50$ & 16.29 & 7.14 \\
\bottomrule
\end{tabular}
\caption{Data of the planetary close encounters for some of the 11 Hungaria fugitives. All the velocities were measured at the distance for an encounter relative to the planet ($2.50\cdot10^{-2}$ AU for the Earth, $1.70\cdot10^{-2}$ AU for Venus and $1.66\cdot10^{-2}$ AU for Mars, as defined in section 2): $\bar{v}_{en}$ is the mean entry velocity, $\bar{v}_{en,max}$ is the mean maximum entry velocity, $\bar{v}_{en,min}$ is the mean minimum entry velocity.  We found that the exit velocity is always equal to the entry velocity inside a range of $0.01km/s$.}
\label{posc}
\end{center}
\end{table}

\begin{table}
\begin{center}
\begin{tabular}{llcccccc}
\toprule
Planet & Ast.    & $\bar{i}_{en}$ & $\bar{i}_{en,max}$ & $\bar{i}_{en,min}$ & $\bar{i}_{ex}$ & $\bar{i}_{ex,max}$ & $\bar{i}_{ex,min}$\\
\midrule
Venus  
       & 1 & $25.80\pm14.42$ & 36.30 & 15.31 & $25.82\pm14.38$ & 36.30 & 15.35\\
\cmidrule{2-8}
       & 2  & $26.16\pm14.43$ & 39.43 & 16.68 & $26.15\pm 14.41$& 39.47 & 16.67\\
\cmidrule{2-8}
       & 3  & $23.94\pm20.60$ & 33.47 & 17.88  & $23.91\pm20.62$ & 33.28 & 17.88\\
\midrule
Earth  
       & 1 & $22.72\pm12.55$  &36.29  & 11.13  & $22.66\pm12.62$  & 36.32  & 11.05\\
\cmidrule{2-8}
       &2  & $25.61\pm11.23$ & 36.85 & 18.92 & $25.51\pm11.26$ & 36.69 & 18.91\\
\cmidrule{2-8}
       &  3  & $27.34\pm22.88$ & 39.25 & 15.72  & $27.39\pm22.96$ & 39.20 & 15.77\\
\midrule
Mars 
       & 1 & $23.26\pm4.14$ & 33.76 & 12.44  & $23.26 \pm 4.14$ & 37.72 & 12.43\\
\cmidrule{2-8}
       & 2 & $27.00\pm4.67$ & 38.59 & 15.66 & $27.00\pm4.68$ & 38.59 & 15.66\\
\cmidrule{2-8}
       &3 & $23.85\pm3.40$ & 32.15 & 16.01 & $23.85\pm3.40$ & 32.15 & 16.02\\
\bottomrule
\end{tabular}
\caption{Data of the planetary close encounters for some of the 11 Hungaria fugitives 
 (1 = (211279) 2002 RN$_{137}$, 2 = (30935) Davasobel and 3 = (39517) 1992 QA.
 All the inclinations was measured at the distance for an
         encounter relative to the planet ($2.50\cdot10^{-2}$ AU for the Earth, $1.70\cdot10^{-2}$ AU for Venus and $1.66\cdot10^{-2}$ AU for Mars, as defined in section 2): $\bar{i}_{en}$ is the mean entry
         inclination, $\bar{i}_{en,max}$ is the mean maximum entry inclination and
         $\bar{i}_{en,min}$ is the mean minimum entry inclination over all clones,
         $\bar{i}_{ex}$ is the mean exit inclination, $\bar{i}_{ex,max}$ is the mean
         maximum exit inclination, $\bar{i}_{ex,min}$ is the mean minimum exit
         inclination.
}
\label{posch}
\end{center}
\end{table}

\noindent

\section{Hungarias as a source of NEAs}

\subsection{Evolution when in NEA orbits}
\label{pastaefagioli}
\begin{figure}
\begin{center}
\includegraphics[width=1.0\textwidth]{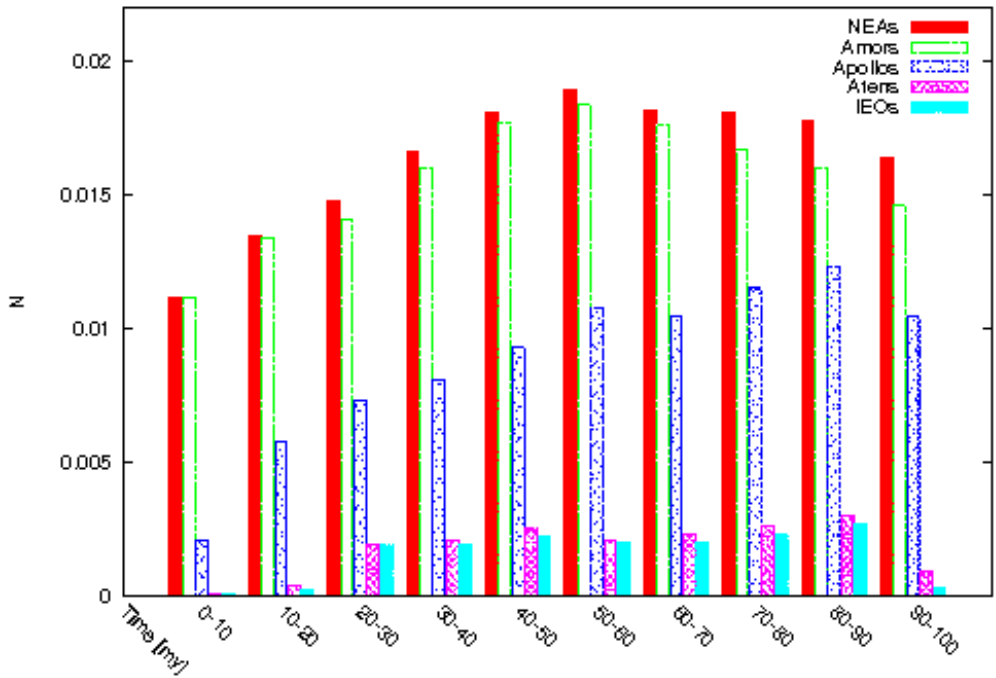}
\caption{Evolution of the Hungaria-NEAs: time versus number of NEAs and their types.
 The number is found like described at the beginnig of Section \ref{pastaefagioli}.
 Note that during this interval of time, the NEA-type can change. In the interval of time, the
first darkest vertical bar on the left is representative of the NEAs and the last, the fourth one, the IEOs.}
\label{evonea}
\end{center}
\end{figure}

Some fugitives start to be NEAs quite soon  with a peak on their rate distribution 
during their orbital evolution, around 60 My,
 as it can be seen clearly from Fig.\ref{evonea}. In this histogram the value $N$ is equal to the
normalized number of asteroids per each time, found with this equation:\\

  $N=\frac{N_{c-n}}{N_c}\cdot \frac{N_f}{N_s}$\\

 $N_{c-n}$, $N_c$, $N_f$ and $N_s$  are respectively the number of Hungaria-clones (HCs)
 which has become NEAs,
 the total number of HCs at the beginning of the orbital evolution, the number
 of the fugitives (Hungarias escaped from the initial sample of 200 real
 Hungarias), 
 and the total number of the initial sample (200),
 considered as representative of the whole family.

The majority of them are Amors, but there is a consistent number of Apollos that stay constantly growing up
till even after all the NEAs reach their maximum value (in fact after the peak of the NEAs, the Amors decrease their 
quantity). Same behaviour, as for the Apollos, is visible for Atens and IEOs; but
 after 90 My their number go down, because of impacts (with the planets or with the Sun).

The average life and the relative release of NEA-types in comparison with some
results of \citet{Bot2002} is described in Table \ref{posc}.

\citet{Bot2002} subdivided the source of NEOs (Near-Earth objects, so not only asteroids,
 but also comets with $q<1.3$ AU) in different regions (\citet{Bot2002} used the so called extended target
 region with these constraints: $q<1.3$ AU, $a<4.2$ AU, $e<1.0$, $i<90^\circ$ and $ 13< H < 22$),
 where we have added the prefix $Neas$:

\begin{itemize}
\item $Neas(\nu_6$), where the secular resonance $\nu_6$ (which occurs 
when the mean precession rates of the longitudes of perihelia of the asteroid and of Saturn are equal to each other) is more active in the Main Belt.
\item $Neas(J3:1)$, where the mean motion resonance with Jupiter J3:1 is more active in the Main Belt.
\item $Neas{IMCs}$, Intermadiate source Mars-crossing asteroids, the subset of the Mars-crossing asteroid  population  that borders the Main Belt.
\item $Neas(OB)$, the Outer Main Belt.
\item $Neas(ECOMs)$, the Ecliptic comets.
\end{itemize}

We find that the average life of the Hungaria-NEAs (Table \ref{posc}, where
 last five rows are referred to  \citet{Bot2002}'s regions) is about the
 double of the asteroid coming from the Outer Main Belt, but much more less
 then the ones coming from the J3:1 and the $\nu6$ resonances
 (see also Table 3 of \citet{Bot2002}). Looking at the average life time spent
 in a region, they are the majorities
 of their life Atens during 100 My of integration, but the maximum life is
 when they are Apollos.

The majority of the fugitives that becomes NEAs are Amors, then IEOs and Atens are moreless in equal number as population of Hungaria-NEAs.
 The percentage for estimating the population is the normalized number of
 asteroids per each time ($N$, described before) in percentage, simply 
\% $=N\times 100$.

Instead, concerning their relative lost of asteroids in the Main Belt, its supply is 
quite comparable to the Outer Main Belt
 Region, but more than 2 times less the  IMCs (see Table \ref{posc}).
 If the lost of these family is constant, in case of no external supply,
 it will finish its own existence in about 3.125 Gy.

\begin{table}
\begin{center}
\begin{tabular}{lccr}
\toprule
Type   & $<L_{TR-100}>$ (My) &$L_{ETR-100,MAX}$& \%  \\
\midrule
IEOs  & 0.052$\pm$0.001  & 3.144 & 0.5 \\
\midrule
Atens & 0.207$\pm$0.001 & 13.981 & 0.5 \\
\midrule
Apollos& 0.101$\pm$0.001   & 82.38 & 2.7  \\
\midrule
Amors  & 0.016$\pm$0.001  &12.893 & 3.2 \\
\midrule
\midrule
Neas  & 0.269$\pm$0.001  & 90.112 & 3.2 \\
\midrule
Source region  & $<L_{TR}>$ (My) &$L_{TR,MAX}$& \%  \\
\midrule
$Neas$ ($\nu_6$) & 6.54 &-&10.2\\
\midrule
$Neas$ ($IMCs$) &3.75 &- &8.1\\
\midrule
$Neas$ ($J3:1$) &2.16 &- &9.3\\
\midrule
$Neas$ ($OB$) & 0.14 &- &4.0\\
\midrule
$Neas$ ($ECOMs$)&45 &- &-\\
\bottomrule
\end{tabular}
\caption{$<L_{TR-100}>$ is the mean life time spent in the target region ($q<1.3$ AU), and $L_{TR-100,MAX}$ is its maximum value, both during an orbital evolution of 100 My. \%  is the percentage of Hungarias (the fraction of the total population of the Hungarias, assumed as the first 200 considered as the sample of the whole family, that enter in the target region) which is source of the relative target region in 100 My. Data that are not about Hungarias (H) comes from \citep{Bot2002} (see Table 3 and Table 1: i.e. for $\nu_6$, initial number or particles were 3519 and 360 of them became NEO on average, so 10.2 \%)  and are put only as comparisons;  here  $<L_{ETR}>$ (My) and $L_{ETR,MAX}$ as the same meaning as previously said with the only different that the value are taken in the extended target region used by \citet{Bot2002} and the objetcs were integrated there for \emph{at least} 100 My.}
\label{posc}
\end{center}
\end{table}

\subsection{Impacts}
\begin{figure}
\begin{center}
\includegraphics[width=1.0\textwidth]{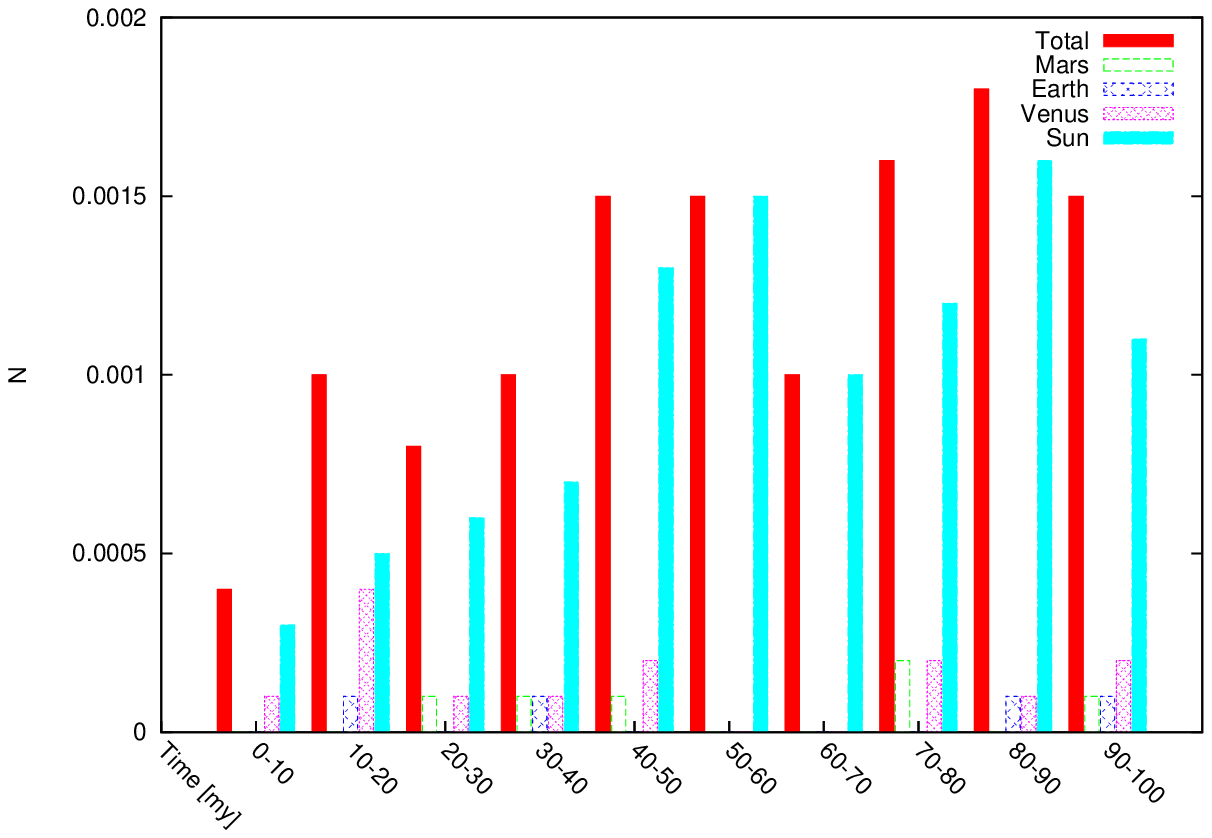}
\caption{Evolution of the Hungaria-NEAs: time versus number of impacts and their types. The impacts with the Sun are more determinant for the final life of the asteroids: the number of impact with the Sun is always major than the number of the impacts with all the planets, even if the impacts with Venus are comparable especially in the first 20 My.}
\label{imp}
\end{center}
\end{figure}

During the integration we found some tens of impacts with the terrestrial planets, Venus captures
 the highest number (Fig.\ref{imp} and Table \ref{imphun}). The highest rate of impacts is between 20 and 30 My, but this
 does not affect the behaviour of increasing the number of NEAs during this interval of time (see Fig.\ref{evonea}).
 Then there is a time span of circa 20 My without impacts with planets (but only with the Sun),
 starting again in the last 30 My. Other objects that finish may have an impact with Jupiter too, but we did not take
in account the study of close encounters with this planet in this work and usually we see a not significant value of
asteroids that cross the Jupiter's orbit.
 So the most important bodies in cleaning up the Hungarias-NEAs seems to be in order the Sun and Venus.
 It is important to notice that there is a certain probability (even if it is very small) that Hungarias
 may hit the Earth and Mars too, that it is important for nowadays studies:
 the probability for these fugitives to hit the Earth and Mars in 100 my respectively is 0.7\% and 1.1\%.   
 Comparing the Hungarias to other sources in respect to the impacts with Venus and Earth,
 the ratio suggests that Hungarias hit more Venus than the others about 3 times more (see Table \ref{Venusimp}),
 even if the statics is minimal and the quantity of impacts are of low numbers.
 The impact flux of Mars is less relevant than Venus in comparison to the Earth.

\begin{table}
\begin{center}
\begin{tabular}{lccccc}
\toprule
Impact flux ratio   & $H$  &$\nu_6$& $3:1$& $IMC$ & $OMB$   \\
\midrule
Venus/Earth         &3.16  & 1.22 &1.19  & 1.10  &0.77 \\
\midrule
Mars/Earth          &0.43  &   -   & -  &  -     & -   \\
\bottomrule
\end{tabular}
\caption{Impact flux ratios (per unit surface area of target), IFR, for Venus/Earth for the individual asteroidal source integrations compared to values in Table 4 of \citet{Gre2012}.}
\label{Venusimp}
\end{center}
\end{table}

The impact-rate is very lower with the average of the planetary crossers, making a comparison with \citet{Iva2002}, 
which take in account the observed planetary crossers with $H<17$, this is clear in Table \ref{imphun}, the highest
contribution by the Hungarias to the impactor populations is $<3\%$. 
Between the impactor population the one in which the Hungarias most contribute is the Martian impactor population.

\begin{table}
\begin{center}
\begin{tabular}{lccc}
\toprule
Planet   & $Hungarias$  & Asteroids ($H<17$)   & \% ratio \\
\midrule
Venus         & 0.014 & 4.500& 0.3\% \\
\midrule
Earth          &0.004  &  3.400& 0.12\%  \\
\midrule
Mars          & 0.006  &0.210& 2.86\% \\
\bottomrule
\end{tabular}
\caption{Average collision probability per one body (planetary crosser) per Gy. The value of the asteroids with $H<17$ 
is from \citet{Iva2002}. \% ratio is the ratio in percentage between the rate of impacts in 1 Gy between the 2 tested
populations.}
\label{imphun}
\end{center}
\end{table}


\subsection{Averaging the elements and analysis}
We averaged all the elements of the 50 clones for each fugitives every 1000 years time step.
We subdivided the clones in two groups: (1) asteroids that become NEAs (\emph{HNeas}) and (2) asteroids
that do not become NEAs (\emph{HNoNeas}).

Finally we consider the data of the group \emph{HNeas} from the time instant when they start to be NEAs to 10 My
 after:
putting time 0 at the moment when they become NEAs, we do the
averaging for the following 10 My (so we average all the Hungaria-NEAs, the clones of the fugitives which are NEAs) and we call this one \emph{NormHNeas}) from now on.
 For all these cases we will show in detail the averaging over 2 fugitives only, representative
for the others 9 ones.

\begin{figure}
\begin{center}
\includegraphics[width=0.45\textwidth]{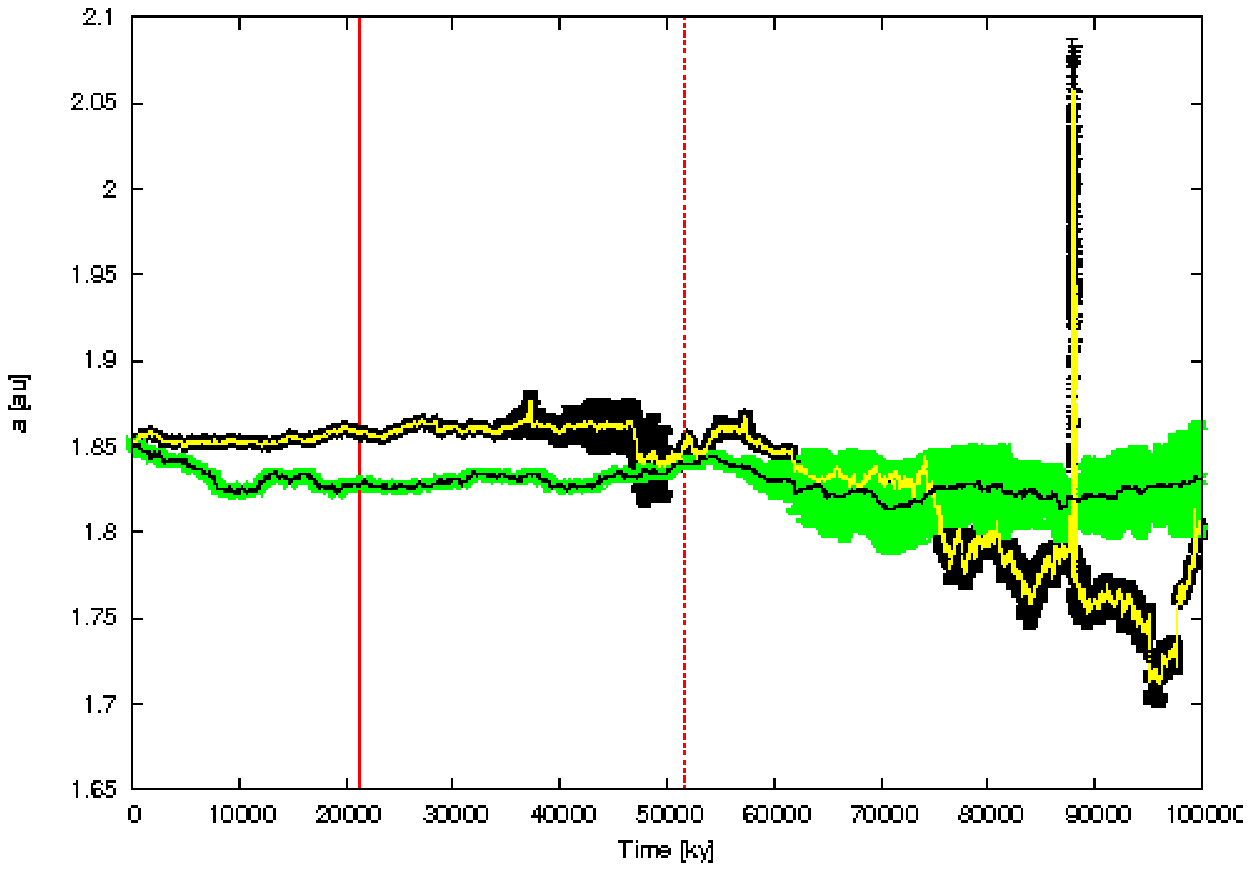}
\includegraphics[width=0.45\textwidth]{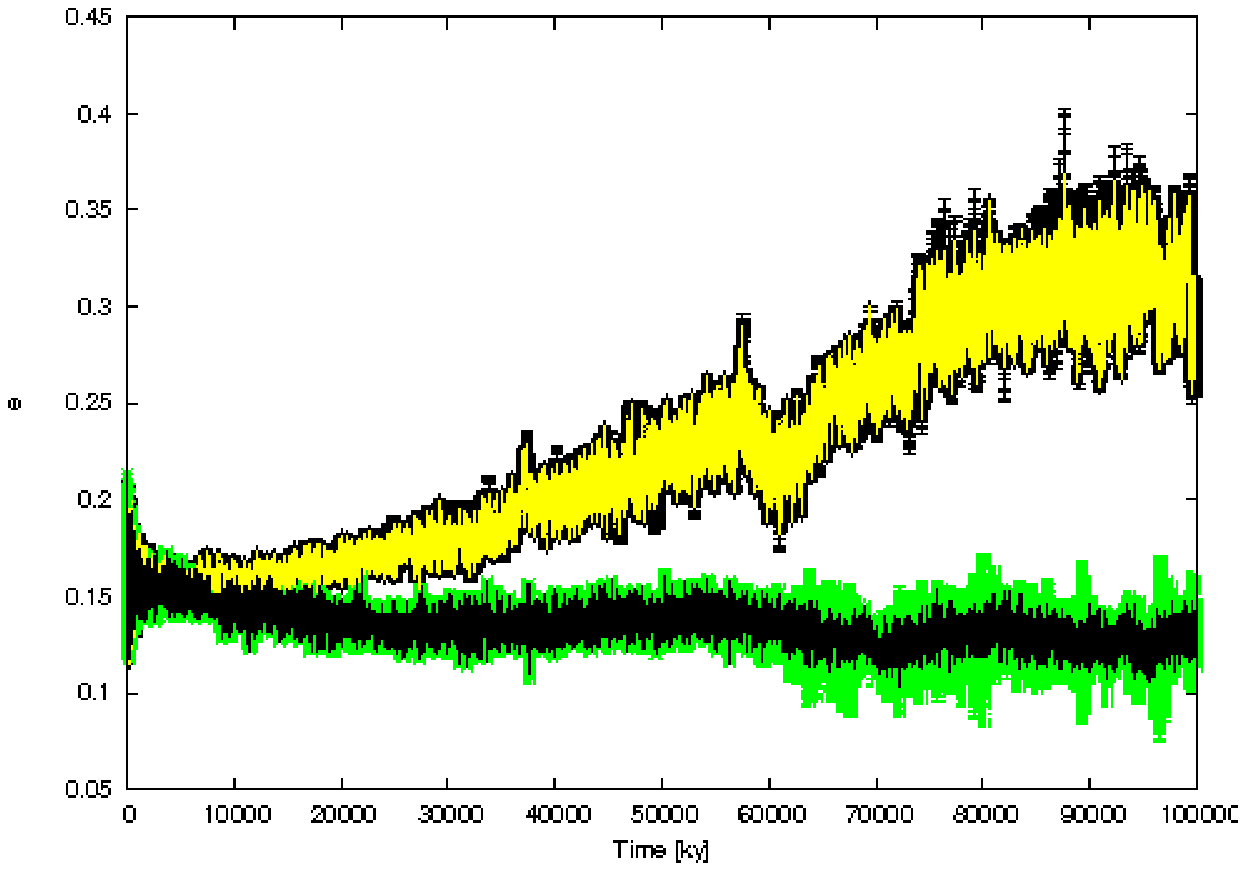}
\caption{Averaging of the elements with the standard deviation of the 50 clones of the asteroid (211279) 2002 RN$_{137}$.
 All points with a standard deviation major than 50\% of the measure of the average has been rejected.
 Left pannel: Semi-major axis versus time for the averaging.
 The first vertical line is the time when the first clone become a NEA.
  The second one is the average time for the clones to become NEAs.
 Right pannel: Eccentricity versus time.}
\label{aveseminea2002}
\end{center}
\end{figure}

In the case of \emph{HNeas} we see (Fig. \ref{aveseminea2002} and \ref{aveseminea1997}) that
 their semi-major axis oscillates much more than \emph{HNoNeas},
 especially after the majorities of them become NEAs and they only decrease this value;
 instead for the \emph{HNoNeas}, the stability is visible (semi-major axis and eccentricity are almost constant)
 till arriving in the proximity of the end of the integration. At the end of evolution, in the last $\sim30$ millions
 year of integration, the scattering increase visibly, due to some impacts and close encounters
with Mars, too.\\
The eccentricity for the \emph{HNeas} have an average increase of about 0.002 $My^{-1}$.

The big scattering visible in Fig.\ref{aveseminea2002} and \ref{aveseminea1997},
 at $\sim$ 88 My for (211279) 2002 RN$_{137}$ and at 65 My for  (152648) 1997 UL$_{20}$,
 is due to some fugitives that escaped immediately out as Centaur or TNO;
 the same is for some bigger oscillations where the scattering is even higher.

The asteroids that do not become NEAs show such a peak in the scattering because they are a small number in comparison to the ones
 that become fugitives.

\begin{figure}
\begin{center}
\includegraphics[width=0.45\textwidth]{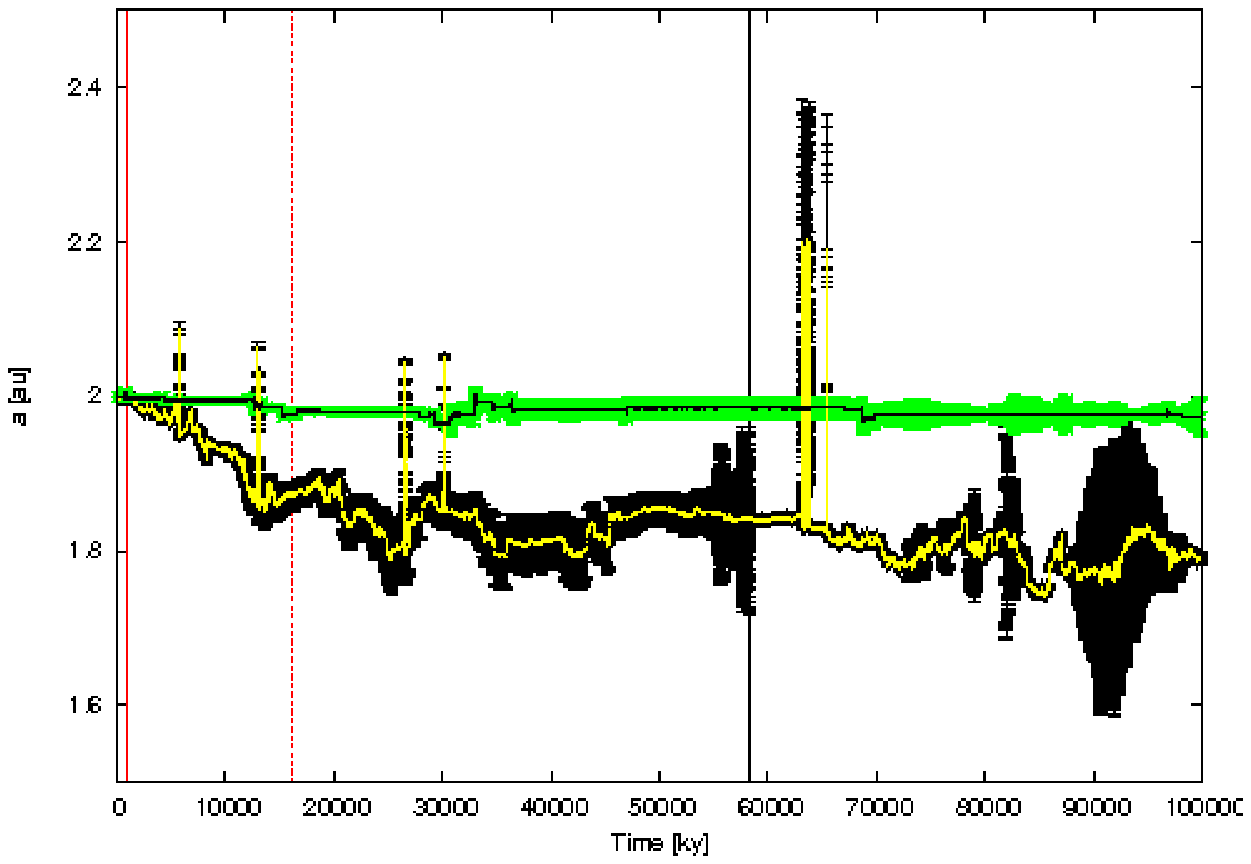}
\includegraphics[width=0.45\textwidth]{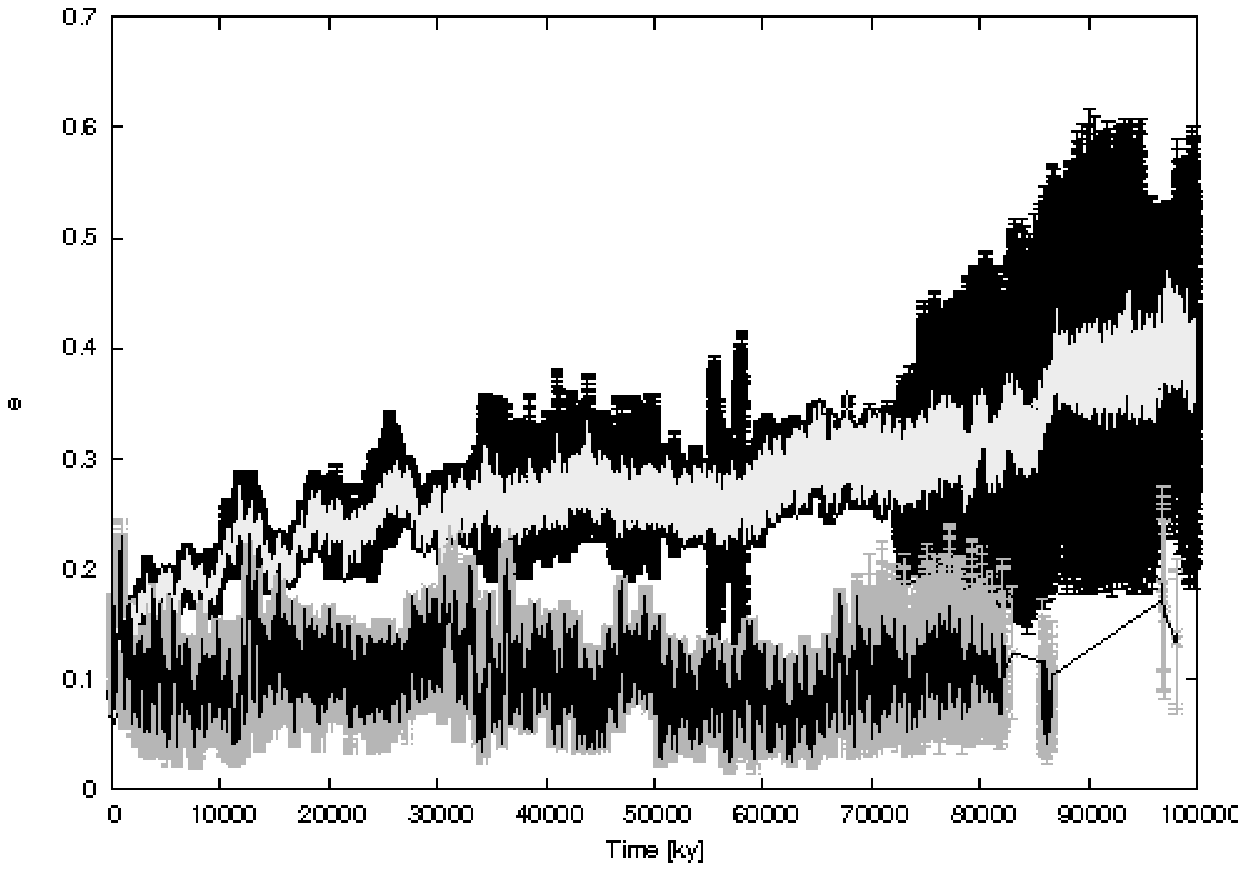}
\caption{Averaging of the elements with the standard deviation of the 50 clones of the asteroid (152648) 1997 UL$_{20}$.
 All points with a standard deviation major than 50\% of the measure of the average has been rejected.
 Left pannel: Semi-major axis versus time for the averaging, at $\sim$ 58 My there is a large scattering. 
 The first vertical line is the time when the first clone become a NEA.
 The second one is the average time for the clone to become NEAs.
 Right pannel: Eccentricity versus time.}
\label{aveseminea1997}
\end{center}
\end{figure}

The  \emph{NormHNeas} initially decreases their semi-major axis more than 0.05 AU (as a whole) in the first million years
 and then after this negative
 gradient, the decreasing of the semi-major axis stops.

Instead the eccentricity after a couple of million years, in the 
cases where the most unstable clones have important close encounters or even impacts, starts to steadily increase, but
below the value of the  \emph{HNeas}.

It is relavant a certain big scattering just when they start to be NEAs. Then some fugitives escaped immediately out as Centaur or TNO, or have some impact and the behaviour become more homogeneous (less scattering).

The analysis on the inclination presents strong scattering giving the studying no possible interpretations.

\begin{figure}
\begin{center}
\includegraphics[width=0.45\textwidth]{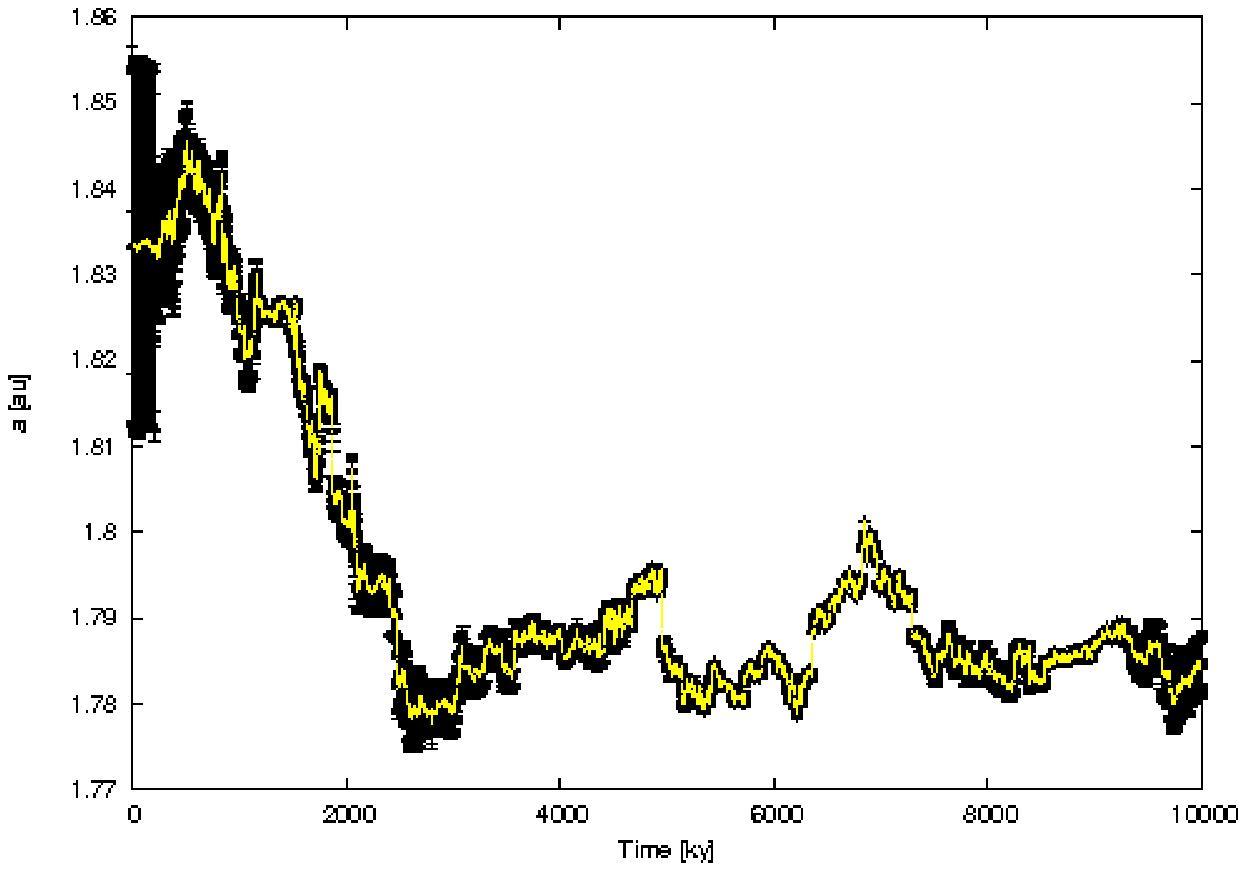}
\includegraphics[width=0.45\textwidth]{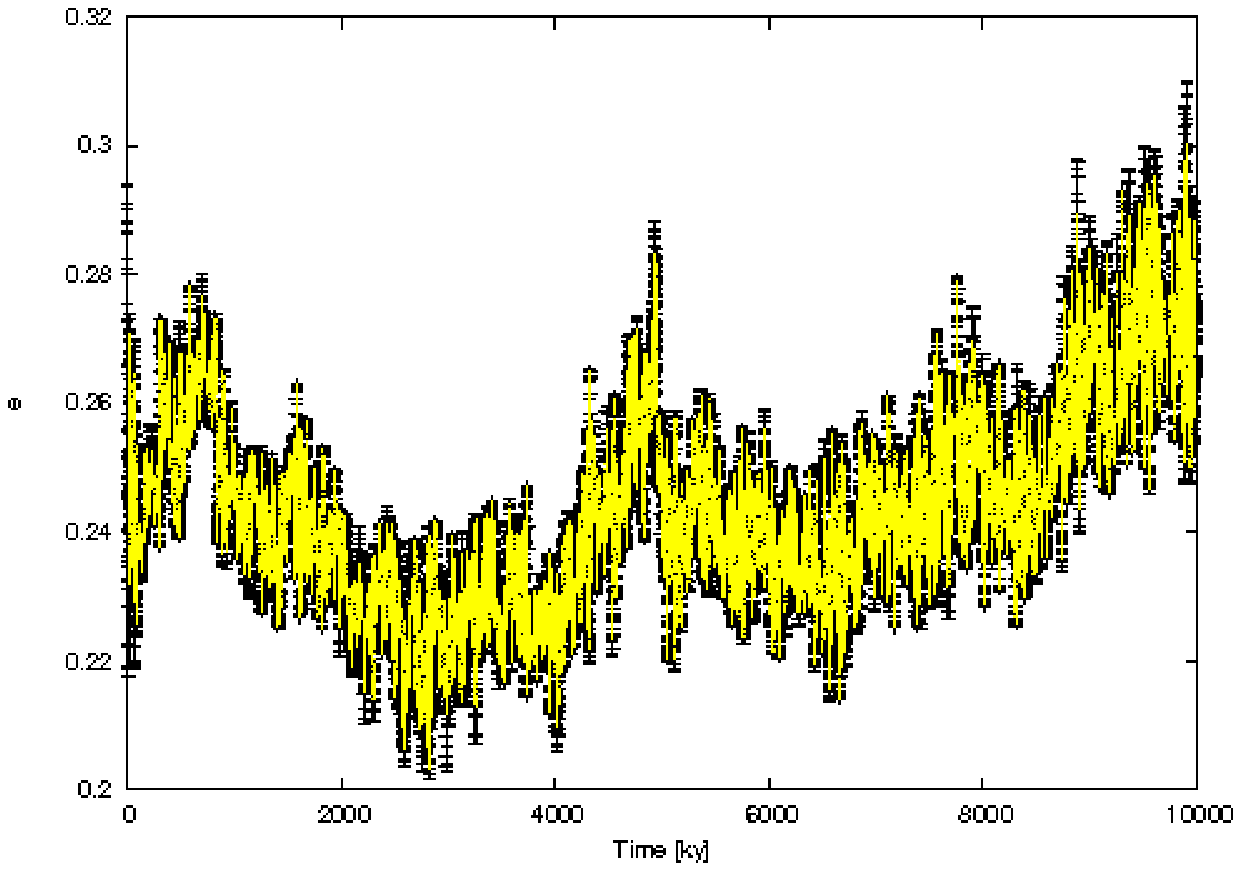}
\caption{Averaging of the elements with the standard deviation of the 50 clones of the asteroid (211279) 2002 RN$_{137}$ when they are NEAs.
 All points with a standard deviation major than 50\% of the measure of the average has been rejected. Left pannel: Semi-major axis versus time for the averaging. Right pannel: Eccentricity versus time.}
\label{aveseminea1997}
\end{center}
\end{figure}

\section{Conclusions}

\label{spec}

As main result of our integration we found that the preponderance of the Hungaria escapers become NEAs, so also Mars-crossers.
  78\% has CE with Mars, 29\% with the Earth and 24\% with Venus. Also 37\% finish their lives as
 Sun-grazers, if they do not collide with a planet before that.

 There is a small number of them (with a very low probability: $<1\%$) that leave the inner main belt crossing the region
 of the Centaurs and Trans-Neptunian Objects (TNOs).

 Some Hungarias  pass a long time in the outer solar system,  via comet-like orbits, because we observed very high
 eccentricities when they are NEAs and even at higher inclinations than the initial conditions.

The average time for a first close encounter with Venus is $\sim 63$ My after the initial conditions,
 but it is somewhat surprising that this time is not much smaller for the Earth $\sim 62$ My;
 the first CE with Mars the average time is lower $\sim 14$ My.
 They become NEAs on average after $\sim$ 47 My.

In the case of Hungaria-derived Apollos or Amors, the Earth is flattening the orbits of the asteroids
 (decreasing the inclination) close to the ecliptic, which is a well known phenomenon.

These asteroids tend to reducing their heliocentric distances, apart some cases that get into a hyperbolic orbit
 (this happens with the narrowest close encounters with Mars after they begin to be Mars-crossers), and some have
 impacts with the Sun, also after only $\sim 10$ My of integration, but the majorities of them after 30 My,
 where the $\nu_6$ secular resonance plays an important role.

During their evolution some escapers end their life with an impact, mostly with the Sun and the rest with the 
planets, in particular the terrestrial planets:  the probability for these fugitives to hit the Earth
 and Mars in 100 My is respectively 0.7\% and 1.1\%. The highest one is for Venus, 2.5\%.

We underline that the majority of the Hungaria escapers change their type-membership when they are NEAs,
 as it has been discussed in detail also by \citet{Mil1989}, \citet{Dvo1999}, and \citet{Dvo2001} and we found
that they have a mean life time equal to 0.27 My on average and their relative probabilities to be source of NEAs
 in 100 My is about 3.2\% with 18\% of these last ones colliding with the Sun.

The maximum release of NEAs by the escapers is  at $\sim$ 60 My after the initial conditions of the integration;
 instead if we look at inside the different NEA-types, we found that Atens and Apollos stop to increase at $\sim$ 90 My.
Their average life-time as NEAs is higher than the one coming from the Outer Main Belt, but the percentage of relative
release is similar; instead this last one is smaller than sources like the $\nu6$, the $J3:1$ and
 in particular more than 2 times less than the  $IMCs$.
 When they are NEAs they pass more
time as Atens in 100 My of evolution after escaping from their proper family.
 If the lost of these family is constant, in case of no external supply,
 it will finish its own existence in about 3.125 Gy.

The majority of the Hungaria fugitives that becomes NEAs are Amors; then IEOs and Atens are more or less in equal number
 as population of Hungaria-NEAs.

Concerning their orbit on average, the semi-major axis continues to decrease, contrary to the eccentricity that is
 continuously growing, $\sim 2\cdot10^{-3} My^{-1}$ with also increasing number of CEs and passing through
 MMRs with different planets and some secular resonances.
CEs are especially with Mars, also many times just before being a NEA.
 We show also that after the Hungarias-NEAs are ``cleaned'' by impacts, the 
remaining ones keep the orbit constant (in semi-major axis and eccentricity) for an
interval time $<$ 10 My like the real ones.

About their initial inclination,
 we think that the former collision started from a period in which massive bodies
 (with diameters bigger than at least 30 km) collided inside the Main Belt region, 
 presumably all started with an initial collision in the inner main belt region for
 a body with probably high inclination ($i\geq23^\circ$).
This collision may have given birth to the E-belt \citep{Bot2011} and this last one
 gave origin to the present Hungarias and a great part of the NEAs.

The fugitives are planet crossing asteroids (PCAs) at least in $91\%$ of the cases in 100 My,
 their angles of deflection can be very high -- rarely up to more than $90^\circ$ 
 (so retrograde orbits, after important CEs), even if on average it is less than $3^\circ$.

 The duration of their close encounters is maximal for Mars, $\sim 0.5$ days, and minimal for Venus,
 $\sim 0.25$ days, somewhere in between for the Earth ($\sim 1/3$ days).
 We could define a relation $2:3:4$ (Venus:Earth:Mars) for the durations of close
 encounters with terrestrial planets, this is because the encounter velocities of the asteroids are very high
 for Venus and lower for Mars.

The Hungarias encounter velocities seem to be faster than the average values for the all the real asteroids that
 come close to the terrestrial planets, especially for the Earth and Venus;
for this reason they probably do look like to cause craters bigger then the aver
at least less than 3\% from the total population of impactors with terrestrial planets,
 but the most important contribution is toward the population which may impact Mars, $\sim2.8$ \%.


\section*{Acknowledgements}

We acknowledge funding from University of Vienna doctoral school IK-1045 and Austrian Science Foundation grant P21821-N19.








\end{document}